\documentclass[aps,twocolumn,pre,showpacs,preprintnumbers,amsmath,amssymb]{revtex4-1}
\usepackage{graphicx}% Include figure files
\usepackage{epsfig}
\input epsf
\epsfclipon
\usepackage{epstopdf}
\usepackage[caption=false]{subfig}

\usepackage[colorlinks]{hyperref}
\mathchardef\mhyphen="2D

\begin{document}
\title{How transfer flights shape structure of the airline network}

\author{Tomasz Ryczkowski}
\author{Agata Fronczak}
\author{Piotr Fronczak}

\affiliation{Faculty of Physics, Warsaw University of Technology, Koszykowa 75, PL-00-662 Warsaw, Poland}

\date{\today}

\begin{abstract}
In this paper we analyze the gravity model in the world passenger air-transport network. We show that in the standard form the model is inadequate to correctly describe the relationship between passenger flows and typical geo-economic variables that characterize connected countries. We propose a model of transfer flights which allows to exploit these discrepancies to discover hidden subflows in the network. We illustrate its usefulness by retrieving the distance coefficient in the gravity model which is one of the determinants of the globalization process. Finally, we discuss the correctness of the presented approach by comparing the distance coefficient to several well known economical events.
 
\end{abstract} 
\maketitle

% 89.75.Da Systems obeying scaling laws
% 89.75.Fb Structures and organization in complex systems
% 89.40.Dd	Air transporation

\section{Introduction} % (fold)
\label{sub:intro}

For many decades, the gravity models have been successfully applied in many different contexts for analyzing socio-economic flows of varying types. The well-known examples include: migration \cite{migration01, migration02, migration03}, consumer spatial behavior \cite{Huff1963}, inter-city telephone communication flows \cite{Krings2009}, hospital-patient flow systems \cite{Lowe1996} and the international trade \cite{Deardorff1998, Anderson1979, Bergstrand1985, Kaski2008, Fagiolo2013, Fronczak2012}.

All these models predict or describe certain behaviors that mimic gravitational interaction as described in Isaac Newton's law of gravity. They assume that a flow between the two places is directly proportional to their importance (expressed in, e.g., population size, gross domestic product (GDP), or some attractiveness index) and is inversely proportional to the physical distance between them. Thus, the simplest form of the gravity equation, written, for example, for the bilateral trade volume, is given by
\begin{equation}
	v_{ij} = G \frac{x_i x_j}{r_{ij}^{\alpha}}
	\label{eq01}
\end{equation}
where $v_{ij}$ is the trade volume between country $i$ and country $j$,  $x_i x_j$ is the product of their GDPs, $r_{ij}$ is the geographic distance between them and $G$ is a constant. Gravity models (GM) work particularly well in the systems where all the places are directly connected (i.e. where the underlying structure is the complete graph). International trade network is a typical example of such a system. The value $v_{ij}$ of products or services exported from the country $i$ to the country $j$ does not affect (at least not directly) the other flows in the network. 

In opposite to the above example, most transport networks involve a series of intermediate stops, which are, themselves, generators of originating and terminating traffic (see e.q. Chapter 7 in \cite{Taaffe1996}). In such networks, especially for large distances, it may happen no direct connection from the location $i$ to the location $j$. In these cases, the potential flow, $f_{ij}^{(g)}$, which might be described by Eq.~(\ref{eq01}), is realized by the increase of subsequent flows $f_{ib_1},f_{b_1b_2},..., f_{b_{n-1}b_n}, f_{b_nj}$. Obviously, this scenario must lead to the observed flow, which differ from the expected one:
\begin{equation}\label{af1}
f_{ij}\neq f_{ij}^{(g)}.
\end{equation}
It means that, in the case of airline networks, the standard gravity model can not be directly used to estimate weights of the existing connection flights.

Contrary to appearances, the divergence of the gravity model with the actual data may prove useful for obtaining deeper insight into the details of the traffic patterns in the transportation networks. In this paper, we demonstrate how one can exploit these discrepancies to discover statistical paths $i-b_1-...-b_n-j$ underlying the observed flows, $f_{ij}$, in the network.

Usually, traffic data are collected in two ways. First, by counting objects (e.g. people, vehicles or information packets) that pass any available link in the network. Such a counting provides one an information about local traffic intensity, however it says nothing about the places the objects started the travel or where they plan to finish. Second, by gathering an information about the origin and destination of each object (e.g. from survey data or from travel tickets) without a knowledge about the detailed path each object follows.

For this study we had at our disposal the data of the first type relating to international flights. We have checked that regardless of the choice of $x_i$ (GDP, population size etc.) in the standard gravity model, the flows $f_{ij}$ are not correctly described by Eq.~(\ref{eq01}). Careful data analysis shows that the observed inconsistency is due to transfer flights, which allow passengers to travel from (or to) less developed regions even though the network is rare. The so-called 'transfer passengers', contribute to reduce flight costs and enhance frequency of flights, which is profitable especially for huge airports. They also have a positive impact on development of small airports. Thus, the understanding of how people choose between different intermediate airports has great practical potential. In this paper, we make a small contribution toward this goal. 

We propose a simple model of connecting flights, which is confirmed by real data. The main assumption of the model is that the potential flows between two countries, $f_{ij}^{(g)}$, which includes all the passengers who start the journey in the country $i$ and end it in $j$, regardless of transfer flights, is given by the gravity law, Eq.~(\ref{eq01}), with $x_ix_j$ standing for the product of GDPs. The mentioned assumption, although can not be directly verified, is well supported by the common observation that the gravity relationship arises from almost any microscopic economic model that includes costs that increase with distance \cite{Deardorff1998}. The last condition is certainly true in most types of transportation networks. 

The final subject of this paper is the discussion of the distance coefficient $\alpha$ in Eq.~(\ref{eq01}). Its behavior over time is strictly related to the globalization process, that can be conceptualized as a continuous reduction of effective distance in the world. Unexpectedly, most studies about gravity models in econometrics clearly show that, since the distance coefficient increases in time, the role of the distance grows simultaneously \cite{2002Coe,2005Brun,2008Disdier,Fronczak2014}. This counter-intuitive result is currently known as the missing globalization puzzle. Here, recovering gravity relationship in the flight network, we are able to analyze the time dependence of the distance coefficient in a typical transportation network.  

In outline, the paper is as follows. First, we describe the data used in this study. Next, we provide a version of the gravity model adapted to the flight network. Then, we introduce the model of connecting flights. Finally, we present the obtained results and discuss the behavior of the distance coefficient.

% subsection intro (end)

\section{Data description} % (fold)
\label{sub:data_description}
Results reported in this paper are based on data provided by International Civil Aviation Organization (ICAO). They contain "annual traffic on-board aircraft on individual flight stages of international scheduled services" \cite{icao}. As a flight stage or a direct flight we understand "the operation of an aircraft from takeoff to landing" \cite{fs}. It means that if a particular flight consists of two (or more) flight stages, we consider it as two (or more) separated direct flights. 

Among many attributes the data contain, such as aircraft type used, the number of flights operated, the aircraft capacity offered and the traffic (passengers, freight and mail) carried, in our analyses we use only the number of passengers traveling between countries. The data are employed to build a sequence of weighted directed networks, $F(t)$,  in the consecutive years t = $1990,...,2011$. In each network, each country is represented by a node and the weight of a link $f_{ij}(t)$ refers to the number of passengers traveling from $i$ to $j$ in year $t$. The flows $f_{ij}(t)$ may vary from a few persons (e.g. $6$ people traveled for Togo to Uruguay in 2004) to several millions of passengers (e.g. $9532303$ people traveled from Great Britain to USA in 2000). 

Apart from traffic data, we also use econometric data from Penn World Table 8.1 \cite{pwt}. To characterize the economic performance of a country we use real GDP at constant 2005 national prices value $x_i(t)$ (in mil. 2005US\$). 
The distance between countries is based on CEPII data \cite{cepii}. Geodesic distances therein are calculated following the great circle formula, which uses latitudes and longitudes of the most important cities/agglomerations (in terms of population). 

% subsection data_description (end)

\section{Simple gravity model} % (fold)
\label{sub:simple_gravity_model}

Before we can verify if the gravity model is able to reproduce the weights of flight connections, we need to determine the value of the constant $G$ in Eq.~(\ref{eq01}). To do it, one has to keep in mind that, in Eq.~(\ref{eq01}), in addition to $G$, there is another free parameter, namely the distance coefficient $\alpha$. This coefficient is usually found from the slope of the linear relation (see, e.g., Fig. 1 in \cite{Fronczak2014})
\begin{equation}
\ln \frac{v_{ij}}{x_ix_j}=\ln G-\alpha \ln r_{ij}.
\end{equation}
We will discuss the distance coefficient in the next Section. At the moment, let us assume, that it its value is known.

In the systems, such as the international trade network, where the flow between $i$ and $j$ only depends on the importance of trading countries, the constant $G$ can be simply obtained from Eq.~(\ref{eq01}),
\begin{equation}
v_{ij}r_{ij}^\alpha=Gx_ix_j,
\end{equation}
after summing over all pairs of countries, i.e.
\begin{equation}
\sum_{i,j}v_{ij}r_{ij}^\alpha=G\sum_{i,j} x_i x_j=GX^2,
\label{gx2}
\end{equation}
where $X$ is a total world GDP and left side of Eq.~(\ref{gx2}) is related to a distance-averaged value of a typical trade channel. This shows that for a fixed value of $\alpha$, the parameter $G$ can be calculated directly from real data. Unfortunately, this is not the case of the airline network. 

\begin{figure*}
	\centering
	\includegraphics[width=\textwidth]{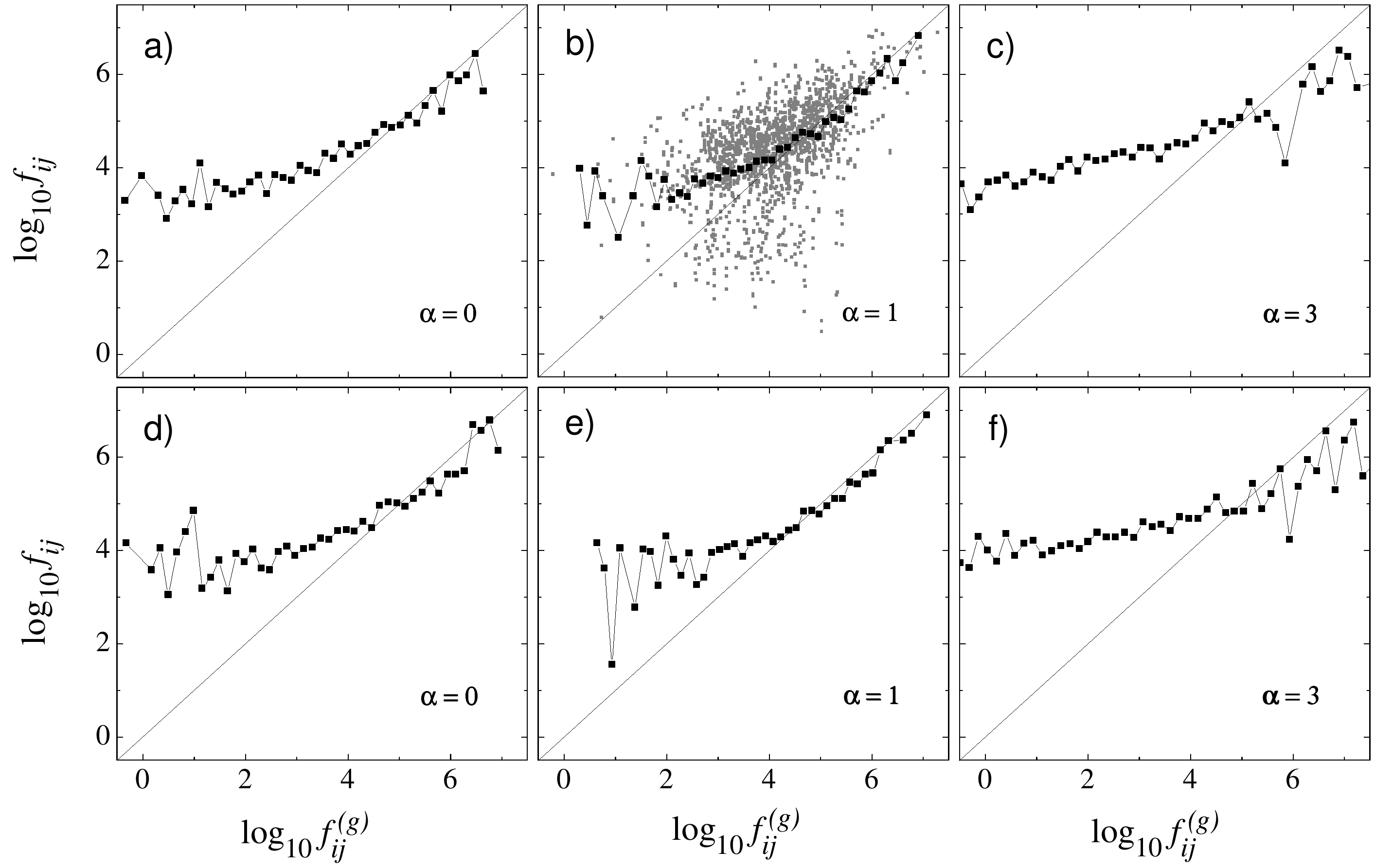}
	\caption{The observed weights of connections in the airline network, $f_{ij}$, vs. their expected values, $f_{ij}^{(g)}$. Plots in the same row correspond to the same year: 1996 (top row) and 2004 (bottom row). Values of the distance coefficient $\alpha$ are indicated in the plots. All data are logarithmically binned (black squares). In panel (b) we have also shown raw data for comparison (gray squares).}
	\label{fig1}
\end{figure*}

In the air-transport network, besides the main contribution to the flow $f_{ij}$ coming from the 'direct passengers' traveling from $i$ to $j$, the value $f_{ij}$ also contains those travelers, for which the flight $i-j$ is only an intermediate link in a longer chain of flights. In other words, the total number of occupied seats, i.e. the sum of all the elements of the matrix $F(t)$,
\begin{equation}\label{af2}
T=\sum_{i,j}f_{ij},
\end{equation}
is larger than the total number of traveling people. In particular, people traveling from $i$ to $j$ with one change occur in this sum twice. Correspondingly, those who travel with two changes (i.e.~with three connecting flights), are taken three times. Therefore, the global traffic $T$ can be estimated in the following way: 
\begin{equation}\label{eq03a}
T\simeq \sum_{l=1}^\infty\;\sum_{(i,j): d_{ij}=l}\hspace{-5pt}l\cdot f_{ij}^{(g)},
\end{equation}
where the summation runs over all pairs of countries $(i,j)$, such that the shortest path between them, in terms of the number of links, is $d_{ij}$, and the expected flow $f_{ij}^{(g)}$ is given by the gravity equation (\ref{eq01}),
\begin{equation}\label{af3}
f_{ij}^{(g)}=G\frac{x_ix_j}{r_{ij}^{\alpha}},
\end{equation}
with $x_ix_j$ standing for the product of GDPs of the connected countries.
It means, that the constant $G$ can be estimated from the following relation
\begin{equation}
G=T\left( \sum_{l=1}^{\infty}\sum_{(i,j): d_{ij}=l} l\cdot \frac{x_ix_j}{r_{ij}^{\alpha}}\right)^{-1}.
\label{eq6aa}	
\end{equation}	
	
Having the constant $G$ estimated, one can plot the observed flows, $f_{ij}$, versus these expected, $f_{ij}^{(q)}$. In Fig.~\ref{fig1}, we present the data for two different years, 1996 and 2004, and for three different values of the distance parameter, $\alpha=0$, $1$, and $3$. The straight line demonstrating the expected flows $f_{ij}^{(g)}$ resulting from Eq.~(\ref{af3}) is also drawn for better comparison. Let us note that the noise, which is inherent to the raw data, makes difficult to clearly estimate the plotted relation (see Fig.~\ref{fig1}b). To overcome this problem, in all the figures we present logarithmically binned data only. 

It is obvious that the direct applicability of the gravity model to the flight network is at least questionable. The best fit is obtained for $\alpha\approx1$ (panels b) and e) in Fig. 1), which coincides with the results obtained by other studies of the distance coefficient in econometric data \cite{Fronczak2014}. However, even if one agree with such a choice of the distance coefficient, the fit is correct only for the right part of each plot. Over a span of at least three decades, the expected, $ f_{ij}^{(g)}$, and the observed flows, $f_{ij}$, differ even by several decades. It seems that there are important factors at play other than economic ones that increase the passenger flow between some countries. In the next section we will show that the connecting flights from the country $i$ to $j$, which do not depend of the economic conditions, $x_ix_j$, of these two countries, can radically change the total flow $f_{ij}$ and explain the discrepancies between the gravity model and real data presented above.

% subsection simple_gravity_model (end)

\section{Model of connecting flights} % (fold)
\label{sub:model_of_connectiong_flights}

We claim that the passenger flow from country $i$ to country $j$, $f_{ij}$, that is observed in data, is composed of the two components:
\begin{itemize}
	\item $f^{(g)}_{ij}$ - the number of passengers traveling directly from the origin of a trip taking place in the country $i$ to the final destination in the country $j$, which, we assume, is given by Eq.~(\ref{af3}),
	\item and the number of passengers, $f^{(transit)}_{ij}$, who use the connection $i \rightarrow j$ as a part of their longer journey.
\end{itemize}

For simplicity we assume that these longer journeys consist of two direct flights only, i.e. we neglect travels with two or more intermediate stops. This assumption seems to be quite strong. For example, in 2004 we have flight data for 151 countries and 22650 possible connections between them. Only 2308 (10\%) of them are direct. There are also 12749 (56\%) shortest paths with the length equal to 2. It means that we take into consideration only 66\% of the all possible connections between the countries. However, it is reasonable to expect that the number of passengers traveling with two or more stops is much less than the lacking 34\% of the global traffic. One of the possible reason for this is that too many transfers complicate a chance for a convenient schedule what costs valuable time. It is usually better to choose then other kind of transportation to reach a destination. We will back to this issue later, when we will discuss the obtained results.

\begin{figure}
	\centering
	\includegraphics[width=\columnwidth]{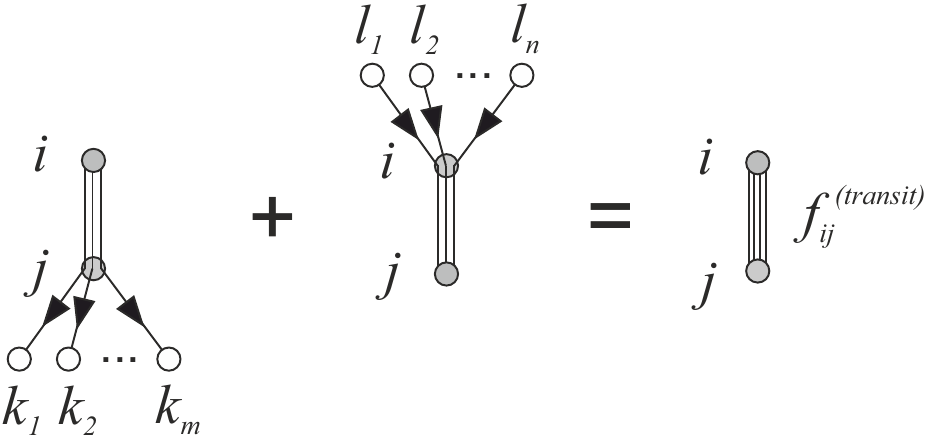}
	\caption{Graphical presentation of the summations in Eq.~(\ref{eq04}).}
	\label{fig2}
\end{figure}

The number of passengers $f^{(transit)}_{ij}$ can be estimated as follows:
\begin{align}\label{eq04}
f_{ij}^{(transit)} = &\sum_{k:i \not \rightarrow k}  f_{ik}^{(g)} \cdot p(i \rightarrow j \rightarrow k)\;+ \\\nonumber &\sum_{l:l \not \rightarrow j}  f_{lj}^{(g)} \cdot p(l \rightarrow i \rightarrow j),
\end{align}
where the first (second) summation is over such nodes $k$ (respectively $l$), that there is no direct connection from $i$ to $k$ (from $l$ to $j$). The term $p(i \rightarrow j \rightarrow k)$ describes the probability that one takes a direct flight from $i$ to $j$ during indirect travel from $i$ to $k$. Contributions of the both summations to the total transit passenger flow $f^{(transit)}_{ij}$ are graphically depicted in Fig. \ref{fig2}. 

The choice of a particular connecting flight from~$i$ through~$j$ to~$k$ (which is expressed by the probability $p(i \rightarrow j \rightarrow k)$) should depend, in a first approximation, on the distance $r_{ij}$ between $i$ and $j$, and the distance $r_{jk}$ between $j$ and $k$. Thus, we omit here the other factors like the convenient flight schedules, type or level of airline service or airport quality, that could influence actual passenger behavior \cite{Johnson2014}. Therefore,
\begin{equation}
		p(i \rightarrow j \rightarrow k) = C \cdot f(r_{ij},r_{jk}), 
\end{equation}
where $C$ is a normalization constant, which is given by
\begin{equation}
	\sum_{j}  p(i \rightarrow j \rightarrow k)=1,
\end{equation}
and the function $f(r_{ij},r_{jk})$ should reflect the tendency of the passengers to choose the shortest, and therefore the cheapest or the fastest connections. Among many possible choices we have chosen the following form for this function
\begin{equation}\label{af4}
f(r_{ik},r_{jk}) = \frac{1}{r_{ij} r_{jk}},
\end{equation}
although the other possible forms, e.g. 
\begin{equation}
f(r_{ik},r_{jk}) = \frac{1}{r_{ij}}+ \frac{1}{r_{jk}},
\end{equation}
lead to similar quantitative results.

Now, having the model defined, one can estimate the total passenger flow between any two countries as the following sum:
\begin{equation}
	f_{ij}^{(mcf)} = f_{ij}^{(g)} + f_{ij}^{(transit)},
	\label{eq05}
\end{equation}
whose components are correspondingly given by Eqs.~(\ref{af3}) and~(\ref{eq04})-(\ref{af4}).

% subsection model_of_connectiong_flights (end)

\section{Results and discussion} % (fold)
\label{sub:results_and_discussion}

\begin{figure}
	\centering
	\includegraphics[width=\columnwidth]{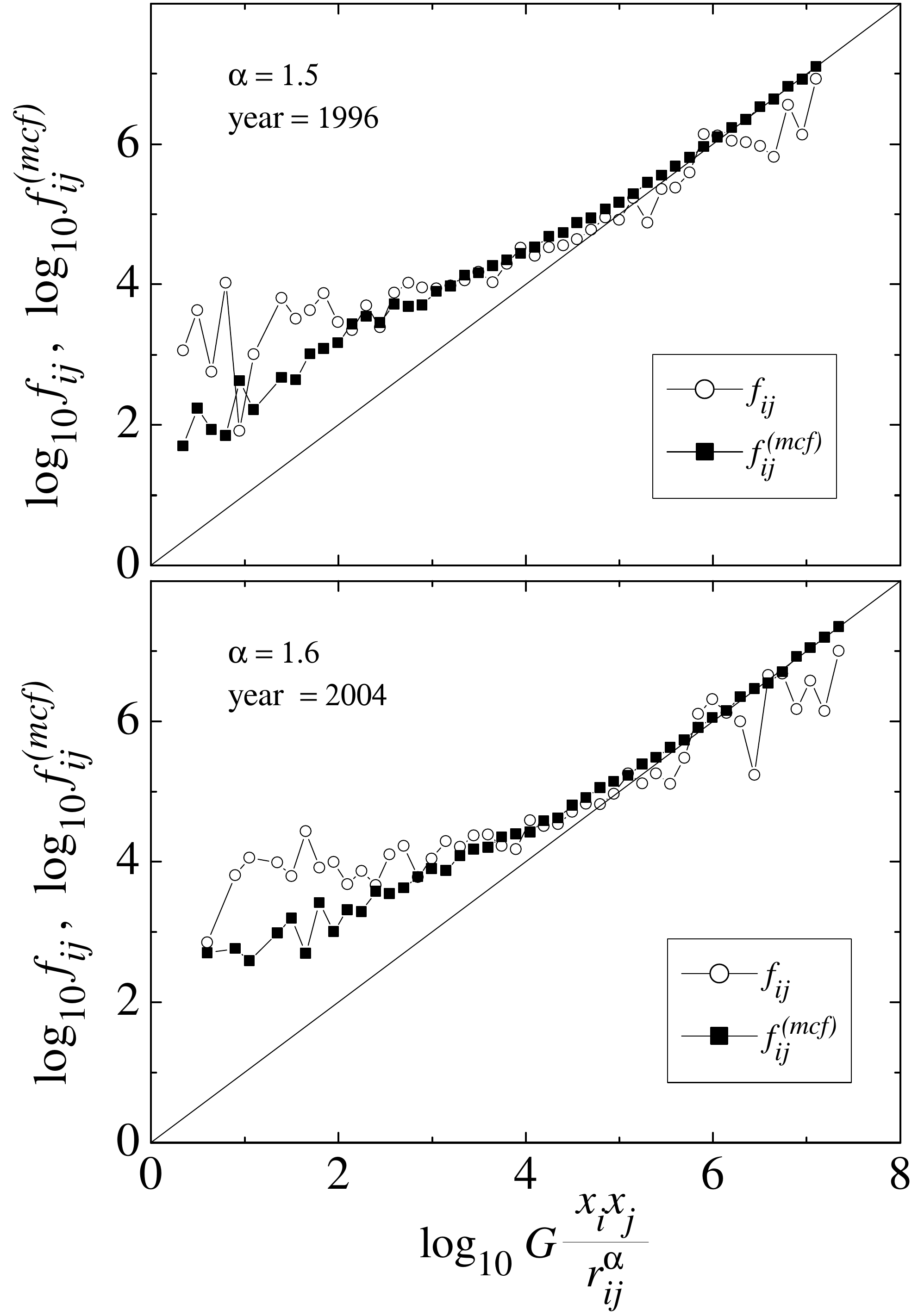}
	\caption{Performance of the model of connected flights (black squares) against real data (open circles) for two years: 1996 and 2004. Straight lines correspond to the standard gravity model.}
	\label{fig3}
\end{figure}

In Fig.~\ref{fig3}, we compare results obtained from our model of connected flights (see Sect.~\ref{sub:model_of_connectiong_flights}) with real data for two different years, 1996 and 2004. We also plot there the straight lines corresponding to the classical GM, Eq.~(\ref{af3}), to demonstrate a significant improvement in performance of the expanded model over GM alone. The largest discrepancies visible in the left part of the plots occur for the long-distance countries with low GDPs, i.e. for large (small) values of the denominator (nominator) in the horizontal axis in Fig.~\ref{fig3}. We have checked that these countries are usually island-based (African, Caribbean and Pacific states) and therefore the travel between them requires multiple transfers - the feature that is not included in our one-stop model. Moreover, a lack of transport alternatives in these countries makes air travel channels more preferred than in the typical continental states. Although it is possible to extend the model to include two-stop connections, we think it is not worth the price, i.e. the significantly increased complexity of the model, especially that its present form correctly predicts more than 98\% of the total passenger flow in the world.

The numerical results for $f_{ij}^{mcf}$ shown in~Fig.~\ref{fig3} have been obtained for the particular values of the distance coefficient $\alpha$ (the reason why we have chosen $\alpha=1.5$ and $\alpha=1.6$ for the years 1996 and 2004 respectively will become clear shortly). One has to keep in mind that the other values of this quantity can lead to the different results and to the better or worse agreement between the model and real data. We can use this observation to select the most probable value of $\alpha$ and to analyze the behavior of the distance coefficient in time. As we mentioned in the introduction, this behavior can be strictly related with the progress of the globalization process in the context of transportation network. Thus, analyzing changes of the distance coefficient would provide another indicator of the rate of the global integration. 

\begin{figure}
	\centering
	\includegraphics[width=\columnwidth]{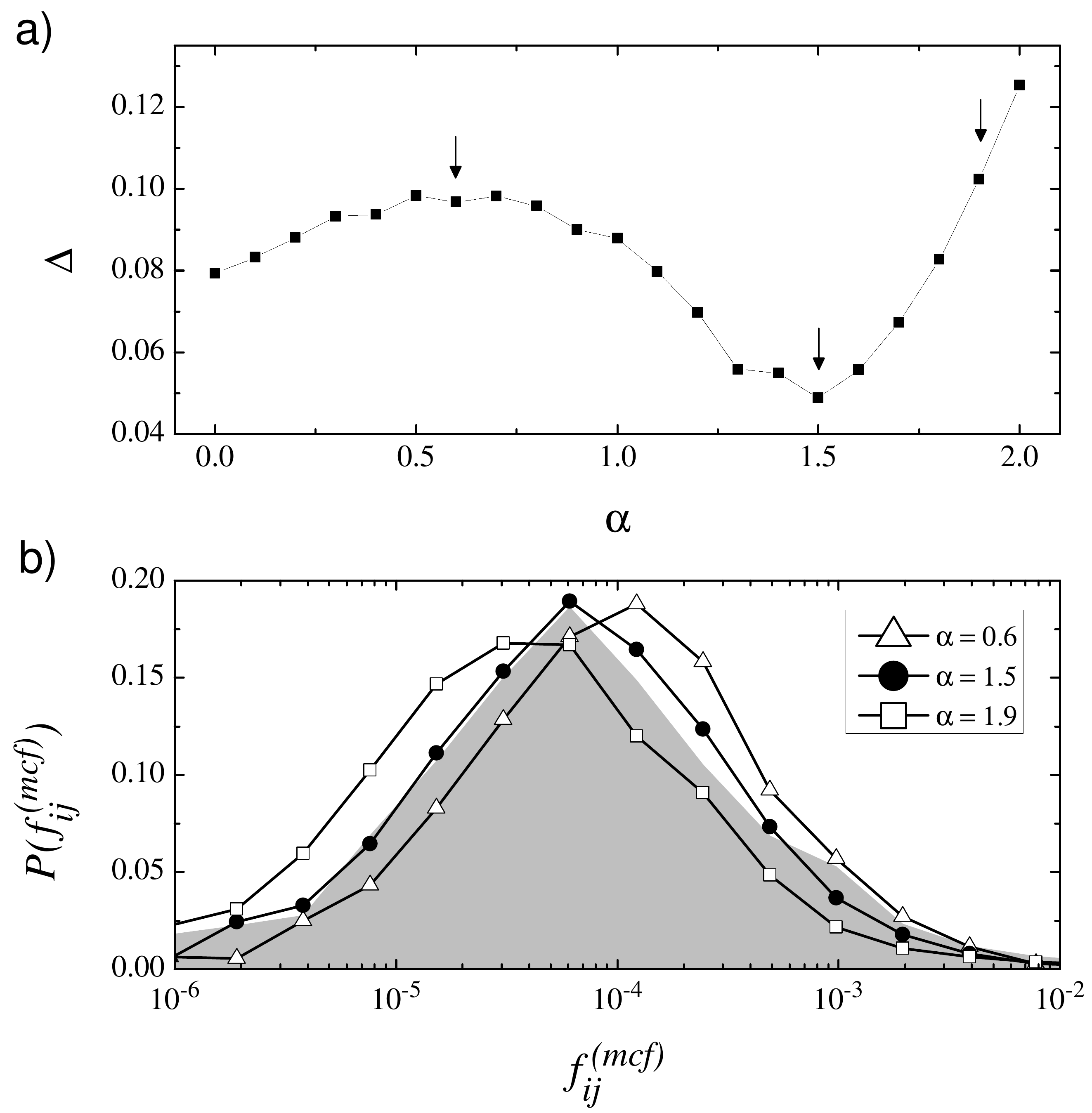}
	\caption{(a) Example of the agreement measure $\Delta({\alpha})$ calculated for different values of the parameter $\alpha$ in the year 1996. The arrows show the values for which three histograms $P(f_{ij}^{(mcf)})(\alpha)$ are shown in panel (b). Gray shaded area represents the histogram $P(f_{ij})$ characterizing real data.}
	\label{fig4}
\end{figure}

For every year in the analyzed period $1990-2011$ we have created the histograms of empirical and modeled flows, $P(f_{ij})$ and $P(f_{ij}^{(mcf)})(\alpha)$ respectively, in $m=15$ logarithmically spaced bins. The examples of such normalized histograms for year 1996 are presented in Fig.~\ref{fig4}b. As one can see, the histograms $P(f_{ij}^{(mcf)})(\alpha)$ created for different values of the $\alpha$ parameter differ in agreement with the histogram of empirical flows (marked by the shaded gray area). To measure this agreement, $\Delta(\alpha)$, we use a simple RMS formula
\begin{equation}
\Delta(\alpha) = \frac{\sqrt{\sum_{h=1}^m{ \left(P_h(f_{ij}) - P_h(f_{ij}^{(mcf)})(\alpha) \right)^2 }}}{N}.
\end{equation} 

In Fig.~\ref{fig4}a we show how this quality measure, $\Delta(\alpha)$, depends on the parameter $\alpha$ in the year 1996. Clearly visible minimum at $\alpha=1.5$ indicates the correct value of the distance coefficient in this year.  

\begin{figure}
	\centering
	\includegraphics[width=\columnwidth]{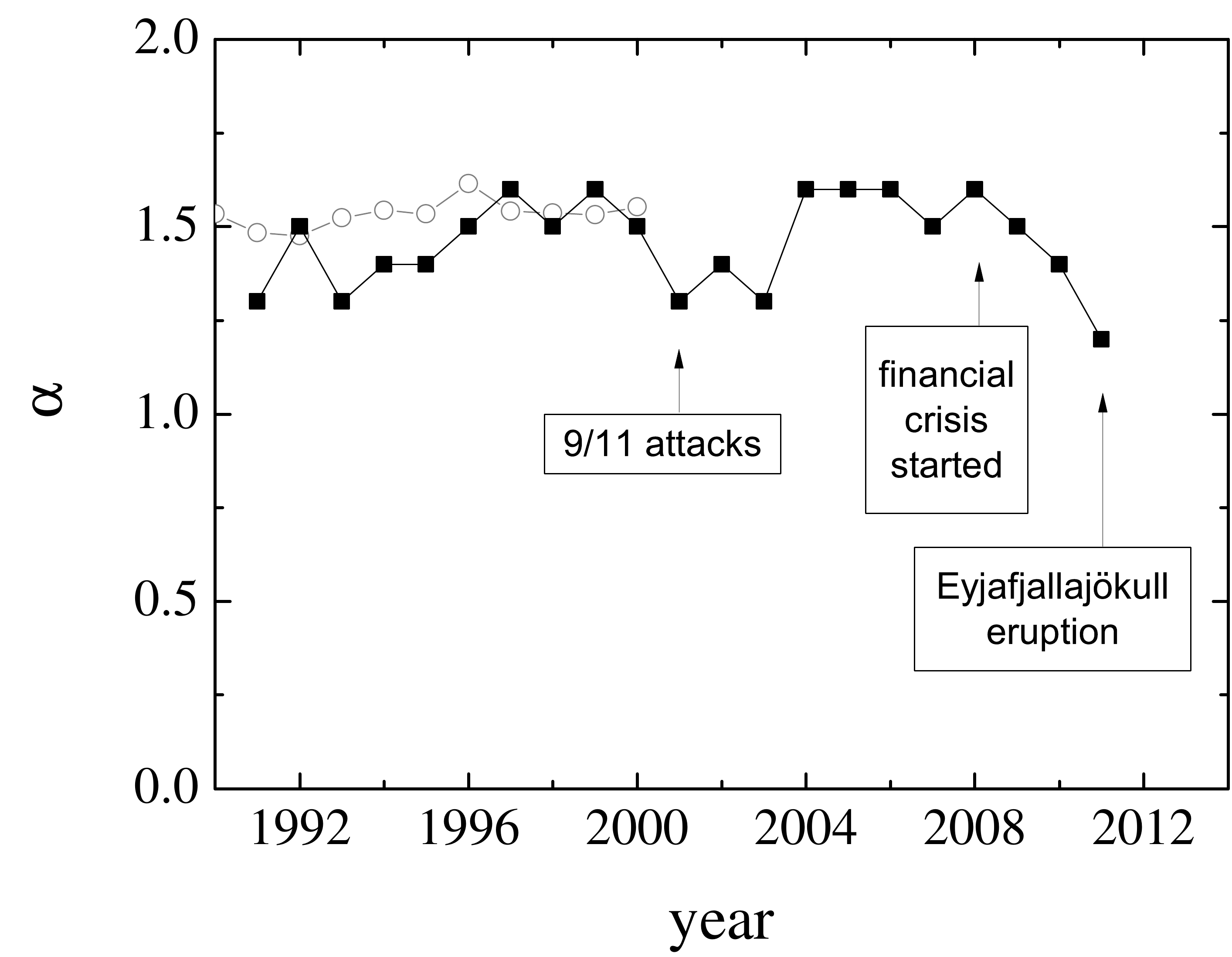}
	\caption{The year-by-year values of the distance coefficient $\alpha$ for the air transportation network resulting from the minimalization of the measure $\Delta(\alpha)$ (black squares) and for the world trade network taken from the paper \cite{Fronczak2014} (gray circles).}
	\label{fig5}
\end{figure}

Figure \ref{fig5} demonstrates the behavior of the distance coefficient for the years $1990-2011$ retrieved by this method. The general conclusion that follows from the figure is that the distance effect in air transportation network is constant over time and the globalization process which is reflected in the distance coefficient has been stabilized in the XXI century. This conclusion confirms the other results (presented by the gray circles in Fig. \ref{fig5}) obtained in \cite{Fronczak2014}, where the authors estimated the distance coefficient for the international trade network. 

Now, let us shortly analyze major fluctuations around this constant distance coefficient. In Fig.~\ref{fig5} we have marked three historical events that could influence the behavior of the distance coefficient in the same way as they had impact on the whole aviation industry. Attacks in New York and Washington D.C. in September of 2001 started a chain of events such as SARS epidemic, additional terrorist attempts, wars, and rising oil prices, that cost the airline industry three years of growth. Airline revenues and traffic surpassed 2000 levels only in 2004 \cite{IATAreport}. The 2008 global financial crisis costed another several years of growth. The effect was further enhanced by the eruption of the Eyjafjallajökull volcano in Iceland in 2010 that caused the closure of airspace over many countries. The correlation between the distance coefficient and all these events visible in Fig. \ref{fig5} confirms that they have a negative impact not only on airline revenues or air traffic but on the whole globalization process.  

Please note, that the globalization process is sometimes conceptualized as a continuous reduction of effective distance in the world \cite{Cairnross1997}, which means that the distance coefficient should vanish in time. However, the observed by us temporary decrease of the distance coefficient is evidently negatively correlated with the progress of globalization. It confirms the recent observations that the distance coefficient is rather associated with the fractal dimension of the considered system and decrease of that coefficient is the effect of decreasing number and weight of air transport connections which reduce dimensionality of the system \cite{Fronczak2014}.

\section{Concluding remarks}

The presented model of connecting flights allowed us to retrieve, from the observed flow between any two countries, the terms corresponding to direct and transfer passengers utilizing this connection. Although we neglected many aspects that influence the choice of intermediate airports by travelers, the model allows to correctly predict more than 98\% of the total passenger flow in the world. The only assumption we had to take into account was that the gravity model is applicable to the case of air transport network. The correctness of the above assumption was confirmed by the time behavior of the retrieved distance coefficient that reflects several historical events with known strong economic impact.

There are still many possible research directions that may be worth exploring in this area. First, the most promising of these seems to be derivation of the so-called fluctuation-response relations \cite{Fronczak2006} that would allow to predict changes in the flows $f_{ij}$ on the basis of changes in GDPs of the connected countries. Now, when we can determine direct and indirect contributions to the particular flow, it should be possible by the analogy to the similar approach done for international trade network \cite{Fronczak2012}. Next, it would be challenging but also rewarding to extend the model taking into account, e.g., time schedules that strongly determine the passenger preference to select a particular intermediate airport. This would allow in general to model microscopic time-dependent flows in the network. Analyzing more detailed level of the air transportation network, in which the nodes represent rather single cities or even airports \cite{Grosche2007} than the whole countries can be also interesting for strategic planning in the airport industry. 

\section*{Acknowledgments}
The work has been supported from the National Science Centre in Poland (grant no. 2012/05/E/ST2/02300).

% subsection results_and_discussion (end)

\end{document}